\documentclass[lettersize,journal]{IEEEtran}
\usepackage{amsmath,amsfonts}
\usepackage[ruled,vlined]{algorithm2e}
\usepackage{algpseudocode}
\usepackage{array}
\usepackage[caption=false,font=normalsize,labelfont=sf,textfont=sf]{subfig}
\usepackage{textcomp}
\usepackage{stfloats}
\usepackage{url}
\usepackage{verbatim}
\usepackage{graphicx}
\usepackage{cite}
\usepackage{multirow}
\usepackage{amssymb}
\usepackage[table]{xcolor}
\usepackage{color , colortbl}
\definecolor{tabhighlight}{rgb}{0.925,1,1}
\usepackage{booktabs} 
\newcommand{\eg}{\emph{e.g.}}

\newcommand{\vs}{\emph{vs.}}

\usepackage{booktabs}
\usepackage{tabularx}
\usepackage{multirow}
\usepackage[table]{xcolor}

\newcolumntype{Y}{>{\centering\arraybackslash}X}
\begin{document}

\title{Multi-Faceted Evaluation and Mitigation of \\ Emotion Hallucinations in MLLMs}

\author{Bowen Zeng, Peipei Song, Weidong Chen, Shengeng Tang, Song Ye, Yuanhong Zhong, Beier Zhu, Xun Yang
\thanks{This work is sponsored by CCF-Lenovo Blue Ocean Research Fund (No 20240106). Corresponding authors: Peipei Song, Beier Zhu and Xun Yang.}
\thanks{Bowen Zeng, Peipei Song, Weidong Chen, Beier Zhu and Xun Yang are with School of Information Science and
Technology, University of Science and Technology of China, Hefei, 230026,
Anhui, China (e-mail: ralph0225@mail.ustc.edu.cn; beta.songpp@gmail.com; chenweidong@ustc.edu.cn; beier.zhu@ustc.edu.cn; xyang21@ustc.edu.cn).}
\thanks{Shengeng Tang is with School of Computer Science and Information Engineering, Hefei University of Technology, Hefei, 230601,
Anhui, China (e-mail: tangsg@hfut.edu.cn). }
\thanks{Song Ye is with Lenovo Group, China.}
\thanks{Yuanhong Zhong is with Chongqing University, Chongqing, 400044, China.}

}



\maketitle

\begin{abstract}
Multimodal large language models (MLLMs) have shown strong potential in open-ended emotion understanding, yet they often generate emotion hallucinations. Evaluating such hallucinations is particularly challenging for two reasons. First, emotion understanding spans multiple cognitive facets, from multimodal perception to psychological reasoning. Second, emotional interpretations are expressed in free-form language, making existing closed-ended protocols insufficient for evaluation. To address these challenges, we introduce the \texttt{EHR} (\texttt{E}motion \texttt{H}allucination \texttt{R}ate) evaluator, which quantifies emotion hallucinations across six facets: \textit{expression}, \textit{action}, \textit{audio}, \textit{instinct}, \textit{logic}, and \textit{conclusion}. Using \texttt{EHR}, we reveal that existing mitigation methods often reduce hallucinations in some facets while aggravating them in others, exposing the limitation of coarse-grained correction and the need for facet-aware localization and mitigation. Motivated by this finding, we propose \texttt{HMER} (\texttt{H}allucination-aware \texttt{M}emory-guided \texttt{E}motion \texttt{R}easoning), a training-free framework for emotion hallucination mitigation. \texttt{HMER} maintains a \textit{Hallucination Memory} that records localized hallucinated claims and enables targeted logit rectification, together with an \textit{Anchor Memory} that preserves reliable intermediate reasoning states to stabilize subsequent generation. By selectively suppressing unreliable cues while preserving trustworthy reasoning context, \texttt{HMER} enables fine-grained mitigation across diverse hallucination facets. Extensive experiments on 19 MLLMs demonstrate the prevalence of emotion hallucinations and the effectiveness of our framework across diverse model architectures.
\end{abstract}


\section{Introduction}
\IEEEPARstart{M}{ultimodal}  emotion understanding seeks to bridge the gap between human emotional intelligence and machine capabilities by enabling models to infer affective states from multimodal signals~\cite{zhang2025mme}. The rise of Multimodal Large Language Models (MLLMs) has shifted multimodal emotion understanding from traditional closed-set classification~\cite{li2017reliable,lian2023mer} toward open-ended, generative interpretation~\cite{lian2025affectgpt, cheng2024emotion, zhao2025r1}.

However, this generative paradigm also introduces emotion hallucinations, where MLLMs produce incorrect or ungrounded emotional interpretations~\cite{bai2024hallucination}. Evaluating such hallucinations requires accounting for two defining characteristics: \textbf{(1) Multi-faceted Cognition:} emotion understanding spans multiple cognitive facets, from multimodal perception to psychological reasoning, beyond the object-centric scope of conventional hallucination evaluation~\cite{rohrbach2018object,li2023evaluating,tu2025ode}. 
For example, as shown in Fig.~1(a), the object-centric CHAIR~\cite{rohrbach2018object} metric cannot assess emotion-specific cues such as \textit{averted gaze} or \textit{aggressive tone}.
\textbf{(2) Free-form Generation:} emotional interpretations are expressed in unconstrained natural language, where different models may rely on different yet valid cues and reasoning trajectories, making closed-ended evaluation protocols insufficient. 
For instance, one model may infer \textit{anger} from a \textit{raised voice}, while another relies on a \textit{frown}. Restricted QA protocols, such as POPE evaluation~\cite{li2023evaluating} shown in Fig.~1(a), cannot capture these distinct yet valid reasoning trajectories.

\begin{figure*}[!t]
    \centering
    \includegraphics[width=\textwidth]{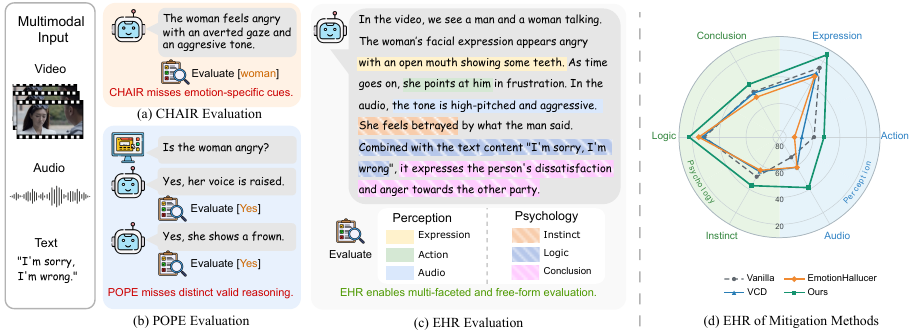}
    \caption{
(a) Object-centric evaluation such as CHAIR miss emotion-specific cues. (b) Restricted QA protocols such as POPE overlook distinct valid reasoning paths in free-form generation. (c) EHR evaluates emotion hallucinations across six facets of multimodal perception and psychological reasoning. (d) \texttt{EHR} further reveals facet-wise differences among mitigation methods, with our method consistently reducing hallucination across all facets.}
    \label{fig:intro}
\end{figure*}

To operationalize these principles, we introduce the \texttt{EHR} (\texttt{E}motion \texttt{H}allucination \texttt{R}ate) evaluator, a psychology-grounded framework for identifying and quantifying emotion hallucinations. Inspired by Ekman's emotion signal systems~\cite{ekman2004emotions}, we define\textbf{ three perception facets}—\textit{expression}, \textit{action}, and \textit{audio}—to assess whether generated interpretations are grounded in observable multimodal cues. Drawing on Shamay-Tsoory's dual-system theory of empathy~\cite{shamay2009two}, we further introduce \textbf{three psychological reasoning facets}—\textit{instinct}, \textit{logic}, and \textit{conclusion}—to characterize higher-level affective inference.
To accommodate free-form outputs, \texttt{EHR} decomposes each generated interpretation into semantically coherent claims, assigns them to the corresponding facets, and verifies each claim against relevant multimodal evidence.  Notably, \texttt{EHR} shows strong agreement with human judgments as shown in Fig~\ref{fig:human_eval}, supporting its reliability.

Our \texttt{EHR} evaluator reveals that existing mitigation methods often improve some facets at the expense of others. As shown in Fig.~\ref{fig:intro}(c), VCD~\cite{leng2024mitigating} and EmotionHallucer~\cite{xing2025emotionhallucer} reduce audio hallucinations while substantially increasing expression and action hallucinations. These trade-offs expose the limitation of current coarse-grained mitigation: suppressing hallucinations in one facet may inadvertently exacerbate errors in others. This motivates the need to localize and correct hallucinated emotional claims in a facet-aware manner. 

\IEEEpubidadjcol
To bridge the gap, we propose \texttt{HMER} (\texttt{H}allucination-aware \texttt{M}emory-guided \texttt{E}motion \texttt{R}easoning), a training-free framework for hallucination mitigation. \texttt{HMER} maintains two memories: the \textit{Hallucination Memory} stores  hallucinated claims with their facet labels and tokens identified by the \texttt{EHR} evaluator, while the \textit{Anchor Memory} preserves trusted reasoning sentences selected by the lowest \texttt{EHR} scores. During decoding, \texttt{HMER} converts the hallucinated tokens stored in the \textit{Hallucination Memory} into a vocabulary-level hallucination direction by taking their vocabulary-wise maximum logits, and suppresses this direction for targeted logit rectification. Meanwhile, sentences in the \textit{Anchor Memory} are fed back as trusted reasoning context, preventing unsupported claims from propagating along the generation trajectory.

Extensive experiments on 19 MLLMs across multiple datasets reveal that emotion hallucination remains pervasive across current MLLMs, cannot be reliably resolved by model scaling or richer modalities, and exhibits a pronounced reliability gap between multimodal perception and psychological reasoning. Our method consistently reduces all facet-wise \texttt{EHR} metrics across all evaluated backbones and datasets, lowering \texttt{EHR}$_{total}$ by 3.27\%--11.85\% while  improving emotion prediction performance.
Our contributions can be summarized as follows:
\begin{itemize}
    \item \textit{\textbf{Evaluation:}} We propose \texttt{EHR} (\texttt{E}motion \texttt{H}allucination \texttt{R}ate) evaluator that decomposes emotion hallucinations into six facets across multimodal perception and affective reasoning, enabling fine-grained evaluation of free-form emotional interpretations.

    \item \textit{\textbf{Methodology:}} We introduce  \texttt{HMER} (\texttt{H}allucination-aware \texttt{M}emory-guided \texttt{E}motion \texttt{R}easoning), which converts localized evaluator feedback into generation-time regulation. \texttt{HMER} maintains a Hallucination Memory for logit rectification and an Anchor Memory for trust-anchored reasoning, thereby reducing both local hallucinations and their propagation through subsequent reasoning.

    \item \textit{\textbf{Performance:}} We systematically evaluate 19 MLLMs spanning video-language, audio-language, omni-modal, and emotion-specialized architectures. The results reveal the prevalence and progression of emotion hallucinations and demonstrate the effectiveness of \texttt{HMER} across representative model backbones.
\end{itemize}

\section{Related Works}
\subsection{Emotion Multimodal Large Language Models}
Multimodal emotion understanding has evolved from traditional discriminative recognition~\cite{li2017reliable,lian2023mer} toward open-ended emotional reasoning~\cite{lian2024open}. Early studies primarily focused on predicting predefined emotion categories from multimodal signals. With the emergence of MLLMs and their enhanced multimodal comprehension and linguistic expressiveness, recent research has shifted toward generative emotional interpretation beyond fixed taxonomies~\cite{zhao2025humanomni,lian2025affectgpt,cheng2024emotion,zhao2025r1,han2025benchmarking, han2026omni, han2026mer}. However, in the absence of explicit grounding constraints, the generative paradigm introduces new hallucination risks, such as fabricated emotional cues and psychologically implausible reasoning~\cite{xing2025emotionhallucer}. Ensuring reliable multimodal emotion reasoning remains an open challenge. 

\subsection{Hallucination in Multimodal Large Models}
Hallucination in multimodal large models has been widely studied in object-centric settings, typically involving errors in object existence, attributes, or relations~\cite{bai2024hallucination}. 
Some studies focus on fine-grained analysis of hallucinations, revealing that errors often stem from inconsistencies between generated semantic claims and visual evidence~\cite{jing2024faithscore}. Evaluation methods quantify such mismatches through discriminative probing~\cite{li2023evaluating}, generative faithfulness metrics~\cite{rohrbach2018object},  claim-level~\cite{jing2024faithscore}, and LLM-based verification~\cite{liu2023mitigating,kaul2024throne}. Mitigation strategies mainly suppress hallucination generation through inference-time intervention~\cite{leng2024mitigating,huang2024opera,jiang2025devils,zhu2026look} or representation-based mitigation~\cite{zhu2026mitigating}, while test-time scaling provides another direction~\cite{stiennon2020learning,wei2022chain,gou2023critic}. Meanwhile, a growing number of dedicated benchmarks~\cite{bang2025hallulens} have been constructed to facilitate systematic and standardized diagnosis of multimodal hallucinations. In contrast, emotion hallucination has only recently gained attention with the rise of emotion MLLMs and exhibits more implicit and multi-faceted characteristics than object hallucination. EmotionHallucer~\cite{xing2025emotionhallucer} represents the first attempt to study emotion hallucination. It uses an adversarial binary question–answer (QA) framework but struggles to capture the diverse errors in open-ended emotional reasoning. In this work, we propose a unified evaluation and-mitigation framework that enables interpretable quantification of emotion hallucinations under a psychology-inspired multi-faceted protocol and converts evaluation signals into controllable generation mechanisms.

\section{Methods}
\begin{figure*}[!t]
  \centering
  \includegraphics[width=\textwidth]{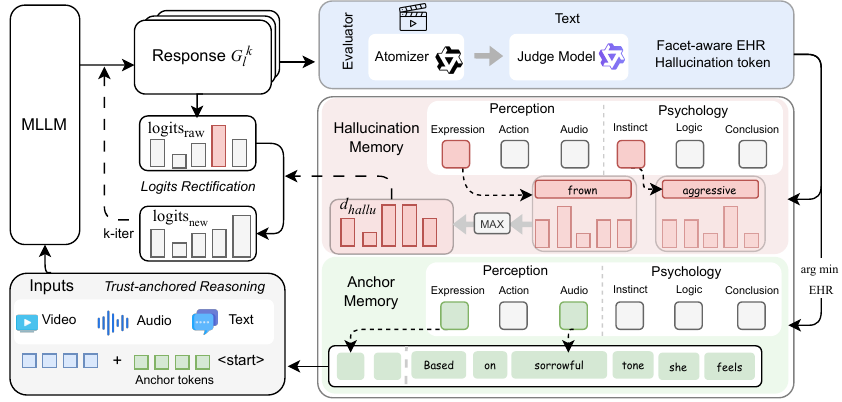}
  \caption{Overview of the \texttt{EHR} evaluator and \texttt{HMER}. The evaluator produces facet-aware hallucination feedback and localized hallucinated tokens from generated responses. \texttt{HMER} then maintains a Hallucination Memory for Logit Rectification, which suppresses a vocabulary-level hallucination direction during iterative decoding, and an Anchor Memory for Trust-Anchored Reasoning, which conditions subsequent generation on the lowest-\texttt{EHR} trusted anchors.
  }
  \label{fig:evaluation}
\end{figure*}

We study emotion hallucination in MLLMs during open-ended emotion reasoning. Given multimodal inputs $X=\{V,A,T\}$, the model generates a natural-language emotional interpretation with $L$ sentences $\{G_l\}_{l=1}^L$ through autoregressive decoding. At each decoding step $t$, the model produces token-level logits ${\bf z}(y_t|X,y_{<t})$ that reflect the internal generation dynamics. First, \texttt{EHR} evaluator defines emotion hallucinations across perception and psychology dimensions and converts free-form generations into claim-level reliability feedback. \texttt{HMER} then uses the identified hallucination spans to construct a hallucination memory for logit rectification, while maintaining a anchor memory to control what can enter subsequent reasoning. An overview is provided in Fig.~\ref{fig:evaluation}.

\subsection{Emotion Hallucination Rate Evaluator}
Unlike object hallucinations that can be verified against explicit entities, emotion hallucinations in open-ended reasoning are obscure and compositional, often manifesting as subtle distortions in multimodal perception or breakdowns in affective inference. To make such failures diagnostically observable, we introduce a structured formulation grounded in psychological theory. Drawing on Ekman's emotion signal systems~\cite{ekman2004emotions} and Shamay-Tsoory's dual-system empathy theory~\cite{shamay2009two}, we conceptualize emotion understanding as a two-stage process of evidence acquisition and cognitive synthesis, and decompose it into six facets organized into two hierarchical levels:
\begin{itemize}
    \item \textbf{Perception Hallucination.} This level measures whether emotional claims are supported by observable multimodal evidence: \textit{\textbf{Expression}} (facial and micro-expression cues), \textit{\textbf{Action}} (body movement and gesture cues), and \textit{\textbf{Audio}} (paralinguistic vocal signals).
    \item \textbf{Psychology Hallucination.} This level evaluates the validity of emotion inference conditioned on perceived evidence: \textit{\textbf{Instinct}} (direct affective appraisal),  \textit{\textbf{Logic}} (evidence-consistent reasoning), and \textit{\textbf{Conclusion}} (globally coherent affective judgment).
\end{itemize}

Given the six-facet taxonomy $\mathcal{S}$ = \{\textit{expression}, \textit{action}, \textit{audio}, \textit{instinct}, \textit{logic}, \textit{conclusion}\}, we construct \texttt{EHR} evaluator to convert a free-form emotional interpretation into localized reliability feedback for metric computation and memory update. As shown in Fig.~\ref{fig:evaluation}, this process consists of claim atomization and verification.

The evaluator first decomposes a generated response $G$ into a set of semantically coherent and independently verifiable claims,
\begin{equation}
\mathcal{C}=\{c_i\}_{i=1}^{N}=\mathrm{Atomize}(G), \quad c_i \mapsto s_i,\ s_i\in\mathcal{S},
\end{equation}
where each claim corresponds to a single perceptual observation or psychology reasoning step, such as a facial movement, an action cue, an acoustic cue, an immediate affective impression, a logical bridge, or a final emotion conclusion. The facet label $s_i$ specifies the evidence required for verification: expression and action claims are checked mainly against visual evidence, audio claims against acoustic evidence, and reasoning claims against multimodal evidence and the stated reasoning context. A judge then returns a binary reliability label,
\begin{equation}
 r_i = \mathrm{Judge}(c_i\mid X,s_i), \quad r_i\in\{0,1\},
\end{equation}
where $r_i=1$ indicates that the claim is supported and $r_i=0$ indicates an emotion hallucination. For generation-time mitigation, the evaluator additionally localizes the token span \(\tau_i\) associated with each hallucinated claim, which is subsequently stored in the Hallucination Memory of \texttt{HMER}.

The facet-wise Emotion Hallucination Rate is then computed as
\begin{equation}
    \texttt{EHR}_s =\frac{1}{|C_s|}\sum_{c \in C_s} \mathbb{I}(r_c = 0), \quad s\in \mathcal{S},
\end{equation}
where $C_s$ denotes the set of claims assigned to facet $s$. Let $\mathcal{S}_{\mathrm{per}}$ and $\mathcal{S}_{\mathrm{psy}}$ denote the perception and reasoning subsets defined above. The two group-level rates are
\begin{equation}
\resizebox{0.96\columnwidth}{!}{$\displaystyle
\texttt{EHR}_{\mathrm{per}}=
\frac{\sum\limits_{s\in\mathcal{S}_{\mathrm{per}}}\!\sum\limits_{c\in C_s}\!\mathbb{I}(r_c=0)}
{\sum\limits_{s\in\mathcal{S}_{\mathrm{per}}}|C_s|},
\quad
\texttt{EHR}_{\mathrm{psy}}=
\frac{\sum\limits_{s\in\mathcal{S}_{\mathrm{psy}}}\!\sum\limits_{c\in C_s}\!\mathbb{I}(r_c=0)}
{\sum\limits_{s\in\mathcal{S}_{\mathrm{psy}}}|C_s|}.
$}
\end{equation}
For an unavailable modality, such as audio in a video-language model, we set the corresponding claim set to $C_s=\varnothing$; it therefore contributes to neither the numerator nor the denominator. The overall rate is
\begin{equation}
\texttt{EHR}_{\mathrm{total}}=
\frac{\displaystyle\sum_{s\in\mathcal{S}}\sum_{c\in C_s}\mathbb{I}(r_c=0)}
{\displaystyle\sum_{s\in\mathcal{S}}|C_s|}.
\end{equation}

Beyond quantifying emotion hallucinations, the \texttt{EHR} scores also serve as reliability criteria for \texttt{HMER}. In particular, \(\texttt{EHR}_{\mathrm{total}}\) is used as the selection metric for constructing the Anchor Memory, where lower scores indicate more reliable reasoning sentences.

\begin{algorithm}[t]
  \small
      \caption{Hallucination-Aware Memory-Guided Emotion Reasoning}
      \label{alg:mem_reasoner}
      \KwIn{multimodal input $X$, maximum rectification round $K$, maximum sentence number $L$}
      \KwOut{anchor archive $\mathcal{M}^A$}
      Init $\mathcal{M}^H \leftarrow \emptyset, \mathcal{M}^A \leftarrow \emptyset$\;
      \For{$l = 1$ \KwTo $L$}{
          $G_l^{0} \leftarrow \mathrm{MLLM}(X, \mathcal{M}^A)$\;
          \For{$k = 0$ \KwTo $K-1$}{
              $(\mathcal{H}_l^k, \texttt{EHR}(G_l^k)) \leftarrow \mathrm{Evaluator}(G_l^k, X, \mathcal{M}^A)$\;
              $\mathcal{M}^H \leftarrow \mathrm{Update}(\mathcal{M}^H, \mathcal{H}_l^k)$\;
              \If{$\texttt{EHR}(G_l^k)=0$}{\textbf{break}}
              $\mathbf{d}_l^k \leftarrow \mathrm{BuildDirection}(\mathcal{H}_l^k)$\;
              $\mathbf{z}_\mathbf{new} \leftarrow \mathrm{Rectify}(\mathbf{z}_\mathbf{raw}, \mathbf{d}_l^k)$\;
              $G_l^{k+1} \sim P(\mathbf{z}_{new})$\;
          }
          $k^* \leftarrow \arg\min_k \texttt{EHR}_{total}(G_l^k)$\;
          $\mathcal{M}^A \leftarrow \mathrm{Update}(\mathcal{M}^A, G_l^{k^*})$\;
          \If{EOS reached}{\textbf{break}}
      }
      \Return $\mathcal{M}^A$\;
\end{algorithm}

\subsection{Hallucination-Aware Memory-Guided Emotion Reasoning}
The evaluator identifies unreliable claims, but post-hoc evaluation cannot prevent them from entering the subsequent reasoning context. This limitation is critical in open-ended emotion understanding, where generation commonly progresses from multimodal observations to psychological interpretation. We propose \texttt{HMER} (\texttt{H}allucination-aware \texttt{M}emory-guided \texttt{E}motion \texttt{R}easoning), a training-free generation-time framework.
\texttt{HMER} performs mitigation by maintaining a \textit{Hallucination Memory} and an \textit{Anchor Memory}, which respectively support Logit Rectification and Trust-Anchored Reasoning. The overall procedure is summarized in Algorithm~\ref{alg:mem_reasoner}.

\noindent\textbf{Memory Construction.}
\texttt{HMER} maintains a Hallucination Memory $\mathcal{M}^H$ and an Anchor Memory $\mathcal{M}^A$. The \textit{Hallucination Memory} stores  hallucinated claims with their facet labels and token spans identified by the \texttt{EHR} evaluator, while the \textit{Anchor Memory} preserves trusted reasoning sentences selected by the lowest \texttt{EHR} scores.

For a candidate $G_l^k$, the evaluator returns the hallucination set $\mathcal{H}_l^k=\{(c_i,s_i,\tau_i)\mid r_i=0\}$, where $c_i$ is an unreliable claim, $s_i$ is its facet, and $\tau_i$ denotes the token corresponding to the claim. If a claim contains multiple hallucinated tokens, each token is stored as a separate entry. \texttt{HMER} writes these entries into the Hallucination Memory,
\begin{equation}
  \mathcal{M}^H = \left\{ (l,k,c_i,s_i,\tau_i,\mathbf{z}_{\tau_i}) \mid r_i=0,\; s_i\in\mathcal{S} \right\},   
\end{equation}
where $l$ and $k$ denote the sentence index and rectification round, respectively, and $\mathbf{z}_{\tau_i}$ contains the vocabulary-level logits of $\tau_i$.


In parallel, the Anchor Memory retains the most reliable candidate generated at each sentence step. Specifically, among all candidates \(\{G_l^k\}_{k=0}^{K}\), \texttt{HMER} selects the one with the lowest total \texttt{EHR} as the trusted anchor:
\begin{equation}
    k^* = \arg\min_{k\in\{0,\dots,K\}} \texttt{EHR}_{total}(G_l^k).
\end{equation}
This selection ensures that the anchor reflects the lowest hallucination risk observed during iterative rectification. The Anchor Memory is then updated as

\begin{equation}
\mathcal{M}_l^A
=
\mathcal{M}_{l-1}^A
\cup
\left\{
G_l^{k^*}
\right\},
\quad
\mathcal{M}_0^A=\varnothing,
\label{eq:anchor_archive}
\end{equation}
where \(\mathcal{M}_l^A\) accumulates trusted reasoning sentences and provides a reliable reasoning context for subsequent generation.

\noindent\textbf{Logit Rectification.}
Given the hallucination feedback stored in the Hallucination Memory from the previous candidate \(G_l^{k-1}\), Logit Rectification retrieves the logits associated with the identified hallucinated spans and constructs a vocabulary-level hallucination direction. For each vocabulary index \(v\), we take the maximum logit across all hallucinated tokens:
\begin{equation}
\mathbf{d}_{l,v}^{k-1}
=
\max_{(c_i,s_i,\tau_i)\in\mathcal{H}_l^{k-1}}
\mathbf{z}_{\tau_i,v}.
\label{eq:hallu_direction}
\end{equation}
The max operator preserves the strongest vocabulary-level signal across hallucinated claim tokens, preventing salient hallucination tendencies from being diluted by weaker ones. Since the direction is constructed from the full vocabulary logits at hallucinated positions, it captures alternative continuations assigned high probability by the model rather than only the tokens selected in the final output. For example, suppressing only ``aggressive'' may cause a lexical substitution such as ``hostile'' without changing the unsupported interpretation, whereas $\mathbf{d}_{l}^{k-1}$ retains the broader distributional tendency associated with the hallucinated content.

We then use this direction to rectify the decoding distribution for the next candidate \(G_l^k\). Let $\mathbf{z}^{(k)}_\mathrm{raw}$ denote the original logits for rectification round $k$, with $\mathbf{p}^{(k)}_\mathbf{raw}=\mathrm{Softmax}(\mathbf{z}^{(k)}_\mathbf{raw})$. The rectified logits are
\begin{equation}
\mathbf{z}^{(k)}_\mathbf{new}
=
\mathbf{z}^{(k)}_\mathbf{raw}
-
\lambda\cdot
\left(
\mathbf{p}^{(k)}_\mathbf{raw}\odot \mathrm{Softmax}(\mathbf{d}_{l}^{k-1})
\right),
\label{eq:logit_suppression}
\end{equation}
where $\lambda$ controls the suppression strength and $\odot$ denotes element-wise multiplication. Applying Softmax to $\mathbf{d}_{l}^{k-1}$ ensures that the suppression term remains non-negative and normalized, thereby making its scale comparable across rectification rounds while preserving the relative strength of hallucination-associated vocabulary entries. Meanwhile, \(\mathbf{p}^{(k)}_{\mathrm{raw}}\) serves as a soft gate, restricting strong suppression to vocabulary entries that are both associated with the hallucination direction and probable under the current decoding context, thereby minimizing interference with unrelated tokens and preserving generation fluency.

Sampling from the rectified distribution produces \(G_l^k\), which is evaluated again to update the Hallucination Memory before the next rectification round. Consequently, \texttt{EHR} evaluator feedback is not only used to rank completed responses, but also directly reshapes the subsequent decoding trajectory.

\begin{table*}[!t]
\centering
\caption{
\textbf{Emotion hallucination rates (\%) across MLLMs}
under different modalities (A: Audio, V: Video, T: Text) on OV-MERD+.
Lower values indicate fewer hallucinations.
The top two results are highlighted in
\colorbox{green!15}{green} (1st) and
\colorbox{blue!15}{blue} (2nd), respectively.
\textsuperscript{$\dagger$}For GPT-4o, we sample 10 frames per video.}
\label{tab:performance_comparison}

\setlength{\tabcolsep}{8.0pt}

\begin{tabular}{l c ccc ccccccccc}

\specialrule{1.2pt}{0pt}{0pt}
\noalign{\vskip 2pt}

\multirow{2}{*}{\textbf{Model}}
& \multirow{2}{*}{\textbf{Size}}
& \multicolumn{3}{c}{\textbf{Modality}}
& \multicolumn{9}{c}{\textbf{Emotion Hallucination Rate (\%)}} \\

\cmidrule(lr){3-5}
\cmidrule(lr){6-14}

& & \textbf{A} & \textbf{V} & \textbf{T}
& \textbf{Expr.}
& \textbf{Act.}
& \textbf{Aud.}
& \textbf{Inst.}
& \textbf{Log.}
& \textbf{Conc.}
& \textbf{Per.}
& \textbf{Psy.}
& \textbf{Total} \\

\midrule

\rowcolor{gray!20}
\multicolumn{14}{c}{\textbf{Open-source MLLMs}\rule{0pt}{6pt}} \\

Qwen2-audio~\cite{chu2024qwen2}
& 7B & $\checkmark$ & & $\checkmark$
& -- & -- & 37.8 & 34.2 & 21.8 & 29.7 & 37.9 & 28.7 & 29.5 \\

SALMONN~\cite{tang2023salmonn}
& 7B & $\checkmark$ & & $\checkmark$
& -- & -- & 38.0 & 30.0 & 24.4 & 31.3 & 36.8 & 28.4 & 29.0 \\

Video-ChatGPT~\cite{maaz2024video}
& 7B & & $\checkmark$ & $\checkmark$
& 28.3 & 44.8 & -- & 35.3 & 18.7 & 41.0 & 41.7 & 31.3 & 35.2 \\

mPLUG-Owl~\cite{ye2023mplug}
& 7B & & $\checkmark$ & $\checkmark$
& 24.7 & 55.0 & -- & 31.8 & 22.3 & 35.2 & 40.3 & 29.6 & 33.2 \\

LLaMA-VID~\cite{li2024llama}
& 7B & & $\checkmark$ & $\checkmark$
& 33.9 & 50.6 & -- & 36.0 & 19.1 & 41.3 & 42.3 & 32.0 & 35.3 \\

Otter~\cite{li2025otter}
& 7B & & $\checkmark$ & $\checkmark$
& 38.3 & 60.9 & -- & 45.6 & 14.3 & 60.2 & 48.5 & 38.7 & 42.2 \\

Chat-UniVi~\cite{jin2024chat}
& 7B & & $\checkmark$ & $\checkmark$
& 19.1 & 33.4 & -- & 35.5 & 22.3 & 32.7 & 24.0 & 29.7 & 27.9 \\

InternVL~\cite{wang2025internvl3}
& 7B & & $\checkmark$ & $\checkmark$
& 7.1 & 14.4 & -- & 24.5 & 19.1 & 15.9 & 9.9 & 19.8 & 17.5 \\

Qwen2.5-VL~\cite{bai2025qwen3}
& 7B & & $\checkmark$ & $\checkmark$
& 6.1 & 13.3 & -- & 24.1 & 18.1 & 16.6 & 9.0 & 19.2 & 16.4 \\

Qwen3-VL-8B~\cite{bai2025qwen3}
& 8B & & $\checkmark$ & $\checkmark$
& \colorbox{blue!15}{4.1}
& \colorbox{green!15}{10.0}
& --
& 22.6
& 18.9
& 13.9
& \colorbox{green!15}{5.7}
& 18.9
& \colorbox{blue!15}{15.3} \\

Qwen3-VL-30B~\cite{bai2025qwen3}
& 30B & & $\checkmark$ & $\checkmark$
& \colorbox{green!15}{3.8}
& 12.1
& --
& \colorbox{blue!15}{19.5}
& 19.2
& 15.8
& \colorbox{blue!15}{6.1}
& 18.3
& \colorbox{green!15}{14.9} \\

R1-Omni~\cite{zhao2025r1}
& 0.5B & $\checkmark$ & $\checkmark$ & $\checkmark$
& 25.4 & 59.2 & 67.6 & 50.9 & 29.0 & 46.0 & 42.2 & 40.6 & 41.3 \\

Emotion-LLaMA~\cite{cheng2024emotion}
& 7B & $\checkmark$ & $\checkmark$ & $\checkmark$
& 40.7 & 77.9 & 40.0 & 50.0
& \colorbox{green!15}{12.4}
& 36.8 & 51.6 & 33.4 & 35.6 \\

HumanOmni~\cite{zhao2025humanomni}
& 7B & $\checkmark$ & $\checkmark$ & $\checkmark$
& 30.1 & 55.7
& \colorbox{green!15}{17.0}
& 22.0
& \colorbox{blue!15}{13.9}
& 33.1 & 39.8 & 23.3 & 32.8 \\

AffectGPT~\cite{lian2025affectgpt}
& 7B & $\checkmark$ & $\checkmark$ & $\checkmark$
& 37.7 & 55.6 & 53.0 & 34.9 & 25.0 & 22.2 & 48.7 & 27.3 & 32.4 \\

Qwen2.5-Omni~\cite{xu2025qwen3}
& 7B & $\checkmark$ & $\checkmark$ & $\checkmark$
& 6.4
& \colorbox{blue!15}{10.8}
& 24.8
& \colorbox{green!15}{19.1}
& 19.0
& \colorbox{blue!15}{12.5}
& 14.0
& \colorbox{green!15}{16.9}
& 16.1 \\

Qwen3-Omni-30B~\cite{xu2025qwen3}
& 30B & $\checkmark$ & $\checkmark$ & $\checkmark$
& 6.5 & 15.5
& \colorbox{blue!15}{23.6}
& 21.9 & 16.5 & 14.9 & 12.9
& \colorbox{blue!15}{18.2}
& 15.9 \\

\midrule

\rowcolor{gray!20}
\multicolumn{14}{c}{\textbf{Closed-source MLLMs}\rule{0pt}{6pt}} \\

GPT-4o\textsuperscript{$\dagger$}~\cite{openaigpt4o}
& -- & & $\checkmark$ & $\checkmark$
& 23.8 & 18.7 & -- & 34.8 & 21.3 & 20.5 & 19.2 & 25.9 & 24.4 \\

Gemini-2.5-flash~\cite{gemini25flash}
& -- & $\checkmark$ & $\checkmark$ & $\checkmark$
& 7.0 & 15.8 & 27.6 & 24.0 & 16.1
& \colorbox{green!15}{12.2}
& 15.4 & 20.1 & 18.0 \\
\specialrule{1.2pt}{0pt}{0pt}

\end{tabular}
\end{table*}
\begin{figure}[t]
  \centering
  \includegraphics[width=\linewidth]{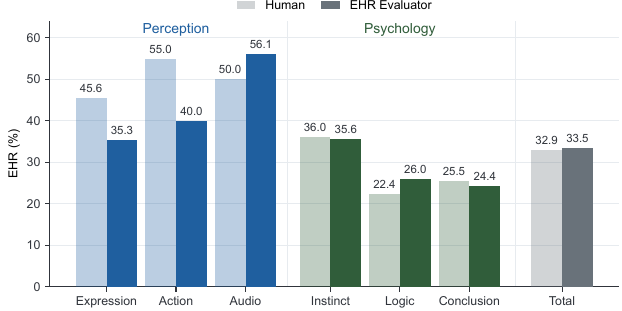}
  \caption{Human \vs  \texttt{EHR} evaluator assessments of emotion hallucination. 
  }
  \label{fig:human_eval}
\end{figure}
\noindent\textbf{Trust-Anchored Reasoning.}
The Anchor Memory accumulates the lowest-\(\texttt{EHR}\) reasoning sentence selected at each generation step, providing a sequence of trusted intermediate anchors. At each sentence generation step, HMER retrieves the Anchor Memory and uses the accumulated anchors as the trusted reasoning context for generating the next sentence:
\begin{equation}
G_{l+1}^{0} = \mathrm{MLLM}\left(X, \mathcal{M}^A\right)
= \mathrm{MLLM}\left(X, G_1^{k^*}, \dots, G_l^{k^*}\right).
\end{equation}

This trust-anchored context prevents rejected candidates from entering the autoregressive history, reducing the inheritance of unsupported perceptual or reasoning claims in subsequent generations. By continually re-anchoring generation to trusted reasoning sentences, Trust-Anchored Reasoning constrains hallucination propagation along the trajectory.

\section{Experiments}
\subsection{Experimental Setup}
\noindent \textbf{Datasets and Metrics.} Experiments are conducted under two settings: {hallucination evaluation} and {hallucination mitigation}. For open-ended emotion reasoning, we conduct zero-shot evaluations on the OV-MERD+~\cite{lian2025affectgpt} dataset. It consists of 532 samples characterized by unfixed emotion categories and a diverse number of labels per instance, providing a fine-grained platform that transcends restricted, predefined taxonomies. For hallucination mitigation, we further evaluate on CMU-MOSI~\cite{chen2017multimodal} and CH-SIMS~\cite{yu2020ch}, covering English and Chinese emotion reasoning respectively, to examine the generalizability of our method across datasets.  We report \texttt{EHR} for reasoning reliability, including \texttt{EHR}$_s$, 
\texttt{EHR}$_\text{per, psy}$, and \texttt{EHR}$_\text{total}$, and F1-Score/Recall~\cite{lian2025affectgpt} for emotion prediction accuracy.

\begin{table*}[!t]
\centering
\setlength{\tabcolsep}{9.0pt}
\fontsize{8}{9}\selectfont

\caption{
\textbf{Performance comparison across datasets and backbones.}
Lower EHR and higher emotion prediction scores are better.
AffectGPT is used as the backbone for cross-dataset evaluation on CMU-MOSI and CH-SIMS.
VCD, ICD, M3ID, and EmotionHallucer are reproduced for fair comparison, and the best results are bolded.
}
\label{tab:overall_results}

\begin{tabular}{lccccccccccc}

\specialrule{1.2pt}{0pt}{0pt}
\noalign{\vskip 2pt}

\multirow{2}{*}{\textbf{Method}}
& \multicolumn{9}{c}{\textbf{Emotion Hallucination Rate ($\downarrow$)}}
& \multicolumn{2}{c}{\textbf{Emotion ACC ($\uparrow$)}} \\

\cmidrule(lr){2-10}
\cmidrule(lr){11-12}

& \textbf{Expr.}
& \textbf{Act.}
& \textbf{Aud.}
& \textbf{Inst.}
& \textbf{Log.}
& \textbf{Conc.}
& \textbf{Per.}
& \textbf{Psy.}
& \textbf{Total}
& \textbf{F1}
& \textbf{Recall} \\

\midrule

\rowcolor{gray!20}
\multicolumn{12}{c}{\textbf{OV-MERD+}\rule{0pt}{6pt}} \\

AffectGPT
& 37.65 & 55.55 & 53.03 & 34.93 & 25.00 & 22.22
& 47.55 & 25.24 & 32.43 & 52.22 & 53.00 \\

\quad + VCD~\cite{leng2024mitigating}
& 37.72 & 50.00 & 49.95 & 31.55 & 26.04 & \textbf{18.40}
& 46.02 & 24.42 & 32.38 & 53.83 & 54.40 \\

\quad + ICD~\cite{wang2024mitigating}
& 25.83 & 54.02 & 41.92 & 35.32 & 20.07 & 23.65
& 37.35 & 24.98 & 29.97 & 53.97 & 55.55 \\

\quad + M3ID~\cite{favero2024multi}
& 35.28 & 48.21 & 45.59 & 30.69 & 20.59 & 18.69
& 42.03 & 22.69 & 30.79 & 53.89 & 53.53 \\

\quad + EmotionHallucer~\cite{xing2025emotionhallucer}
& 29.49 & 66.66 & 50.51 & 33.87 & 22.60 & 26.60
& 48.54 & 29.86 & 31.70 & 47.11 & 49.34 \\

\rowcolor{tabhighlight}
\quad \textbf{+ Ours}
& \textbf{13.81}
& \textbf{39.31}
& \textbf{34.54}
& \textbf{25.70}
& \textbf{19.80}
& 21.17
& \textbf{28.14}
& \textbf{22.50}
& \textbf{23.52}
& \textbf{55.23}
& \textbf{58.45} \\

\midrule

Qwen2.5-Omni
& 6.45 & 10.89 & 24.80 & 19.11 & 19.01 & 12.55
& 14.02 & 16.97 & 16.10 & 57.32 & 55.68 \\

\quad + VCD~\cite{leng2024mitigating}
& 8.15 & 12.76 & 25.22 & 19.44 & 15.50 & 12.16
& 15.50 & 15.85 & 15.74 & 56.78 & 55.83 \\

\quad + ICD~\cite{wang2024mitigating}
& 5.01 & 10.14 & 29.17 & 15.88 & 13.23 & 16.22
& 13.88 & 15.07 & 14.60 & 56.24 & 55.21 \\

\quad + M3ID~\cite{favero2024multi}
& 4.44 & 9.90 & 19.92 & 17.80 & 12.76 & 12.78
& 9.17 & 14.30 & 12.84 & 55.02 & 53.38 \\

\quad + EmotionHallucer~\cite{xing2025emotionhallucer}
& 6.21 & 12.96 & 18.08 & 15.83 & \textbf{11.30} & 13.11
& 10.74 & \textbf{13.98} & \textbf{12.14} & \textbf{58.61} & 54.53 \\

\rowcolor{tabhighlight}
\quad \textbf{+ Ours}
& \textbf{4.31}
& \textbf{9.84}
& \textbf{15.25}
& \textbf{12.84}
& 18.54
& \textbf{11.93}
& \textbf{9.14}
& 14.28
& 12.83
& 58.02
& \textbf{57.23} \\

\midrule

R1-Omni
& 25.38 & 59.23 & 67.61 & 50.89 & 28.99 & 46.01
& 42.16 & 40.62 & 41.28 & 42.88 & 40.55 \\

\quad + VCD~\cite{leng2024mitigating}
& 30.17 & 68.40 & 59.03 & 55.34 & 27.43 & 46.94
& 48.71 & 41.70 & 44.72 & 42.22 & \textbf{41.49} \\

\quad + ICD~\cite{wang2024mitigating}
& 27.59 & 71.76 & 51.47 & 51.64 & 22.05 & 47.66
& 44.67 & 37.57 & 41.24 & 39.94 & 38.29 \\

\quad + M3ID~\cite{favero2024multi}
& 24.82 & 65.94 & 60.17 & 46.82 & 24.59 & 46.85
& 43.38 & 36.82 & 40.62 & 36.72 & 33.68 \\

\quad + EmotionHallucer~\cite{xing2025emotionhallucer}
& 32.77 & 73.84 & 59.00 & 56.43 & 24.76 & 50.62
& 47.71 & 44.52 & 45.66 & 38.76 & 39.68 \\

\rowcolor{tabhighlight}
\quad \textbf{+ Ours}
& \textbf{14.09}
& \textbf{51.84}
& \textbf{41.49}
& \textbf{43.18}
& \textbf{17.49}
& \textbf{39.55}
& \textbf{30.10}
& \textbf{32.70}
& \textbf{31.47}
& \textbf{44.24}
& \textbf{41.49} \\

\midrule

\rowcolor{gray!20}
\multicolumn{12}{c}{\textbf{CMU-MOSI}\rule{0pt}{6pt}} \\

AffectGPT
& 38.12 & 50.28 & 80.35 & 46.84 & 23.65 & 59.94
& 65.69 & 42.49 & 52.87 & 77.77 & 77.94 \\

\quad + VCD~\cite{leng2024mitigating}
& 40.00 & 48.67 & \textbf{71.30} & 46.30 & 23.02 & 60.07
& 60.31 & 42.36 & 50.50 & 75.73 & 76.19 \\

\quad + ICD~\cite{wang2024mitigating}
& 36.48 & 44.19 & 80.33 & 44.62 & 17.39 & 62.79
& 63.98 & 40.89 & 50.74 & 77.84 & 78.05 \\

\quad + M3ID~\cite{favero2024multi}
& 44.61 & 56.25 & 78.09 & 47.04 & 21.41 & 60.34
& 65.79 & 42.02 & 52.47 & 75.84 & 76.37 \\

\quad + EmotionHallucer~\cite{xing2025emotionhallucer}
& 28.32 & 45.93 & 77.08 & 54.43 & 17.28 & 69.98
& 58.84 & 45.30 & 50.73 & 73.65 & 72.76 \\

\rowcolor{tabhighlight}
\quad \textbf{+ Ours}
& \textbf{19.83}
& \textbf{32.40}
& 75.43
& \textbf{39.87}
& \textbf{13.45}
& \textbf{59.04}
& \textbf{52.96}
& \textbf{36.97}
& \textbf{42.98}
& \textbf{78.41}
& \textbf{78.57} \\

\midrule

\rowcolor{gray!20}
\multicolumn{12}{c}{\textbf{CH-SIMS}\rule{0pt}{6pt}} \\

AffectGPT
& 31.48 & 51.85 & 56.14 & 33.07 & 24.54 & 21.18
& 47.96 & 24.54 & 32.20 & 79.99 & 80.41 \\

\quad + VCD~\cite{leng2024mitigating}
& 37.97 & 65.82 & 48.58 & 36.71 & 20.09 & \textbf{19.19}
& 46.37 & 23.99 & 33.68 & 75.71 & 76.55 \\

\quad + ICD~\cite{wang2024mitigating}
& 30.22 & 52.54 & 46.71 & 35.58 & \textbf{13.72} & 23.22
& 41.09 & 22.55 & 30.25 & 76.78 & 77.32 \\

\quad + M3ID~\cite{favero2024multi}
& 33.62 & 50.00 & 51.21 & 34.24 & 24.89 & 20.34
& 44.76 & 25.88 & 33.31 & 78.32 & 79.12 \\

\quad + EmotionHallucer~\cite{xing2025emotionhallucer}
& 30.03 & 63.68 & 50.50 & 48.89 & 15.79 & 32.38
& 44.66 & 28.13 & 34.81 & 71.10 & 73.45 \\

\rowcolor{tabhighlight}
\quad \textbf{+ Ours}
& \textbf{13.29}
& \textbf{22.26}
& \textbf{24.71}
& \textbf{19.12}
& 21.75
& 20.98
& \textbf{18.84}
& \textbf{20.93}
& \textbf{20.35}
& \textbf{82.41}
& \textbf{82.47} \\

\specialrule{1.2pt}{0pt}{0pt}

\end{tabular}
\end{table*}
\noindent \textbf{Evaluated MLLMs.} To analyze emotion hallucinations, we evaluate both open-source and closed-source MLLMs spanning different capability profiles, including emotion-specialized (\eg, AffectGPT~\cite{lian2025affectgpt}, R1-Omni~\cite{zhao2025r1}), audio-centric (\eg, Qwen2-Audio~\cite{chu2024qwen2}), video-understanding (\eg, Video-ChatGPT~\cite{maaz2024video}), general-purpose models (\eg, Qwen2.5-Omni~\cite{xu2025qwen3}), and so on. Experiments are conducted via official implementations or APIs. 
For hallucination mitigation experiments, we select AffectGPT, Qwen2.5-Omni, and R1-Omni as backbones, and compare our method with representative mitigation solutions: VCD~\cite{leng2024mitigating}, ICD~\cite{wang2024mitigating}, M3ID~\cite{favero2024multi} (contrastive decoding for object hallucination), and EmotionHallucer~\cite{xing2025emotionhallucer} (self-verification for emotion hallucination). 

\noindent \textbf{Implementation Details.} For hallucination evaluation, we employ Qwen3~\cite{yang2025qwen3} as our atomizer, and include Qwen2.5-Omni~\cite{xu2025qwen3} and Qwen3 as the judge models. The models are locally deployed. 
Regarding the iterative rectification, we set the logit suppression coefficient $\lambda$=50, maximum retry iterations $K$=3, and maximum sentence number $L$=20. All experiments are conducted on an A800 GPU.

\subsection{Emotion Hallucination Evaluation}
\label{sec:hall_eval}
\noindent\textbf{Benchmarking MLLMs.}
Table~\ref{tab:performance_comparison} presents a comprehensive evaluation of state-of-the-art MLLMs under the proposed EHR protocol. 
We have three key observations. (1) Emotion hallucination is a pervasive issue in existing MLLMs. Across models, the overall \texttt{EHR}$_{total}$ remains substantial (\eg, Otter: 42.2\%, LLaMA-VID: 35.3\%).
(2) Model scaling cannot resolve hallucination; for instance, Qwen3-VL-30B lowers \texttt{EHR}$_{total}$ relative to its 8B counterpart, yet residual errors persist across multiple facets. Also, enriching modality is insufficient; omni-models do not consistently outperform VL models (\eg, Qwen3-Omni-30B \vs Qwen3-VL-30B), suggesting potential multimodal interference. 
(3) A pronounced discrepancy exists between perception and reasoning reliability. For example, Qwen3-VL-8B has superior perceptual grounding (\texttt{EHR}$_{per}$=5.7\% ), but its reasoning hallucination remains elevated (\texttt{EHR}$_{psy}$=18.9\%), indicating that emotion hallucination is inherently a multi-level reliability failure.

\noindent\textbf{Human Evaluation.}
To validate the evaluator, we perform a human study on 100 randomly sampled instances. Three expert annotators score the model outputs across six fine-grained emotion hallucination dimensions. As shown in Fig.~\ref{fig:human_eval}, the automatically computed \texttt{EHR} closely matches human judgments. Specifically, the Pearson correlation~\cite{pearson1896vii} between the two sets of scores is 0.8364, the Spearman rank correlation coefficient~\cite{spearman1961proof} is 0.9167, the Cohen Kappa coefficient~\cite{mchugh2012interrater} is 0.7692, and the intra-class correlation coefficient (ICC)~\cite{koch2004intraclass} is 0.7959. These results demonstrate that the evaluator provides a scalable proxy for human assessment.

\begin{table*}[t]
\centering
\setlength{\tabcolsep}{10.0pt}

\caption{
\textbf{Ablation study and comparison with test-time scaling.}
We evaluate the contributions of Logit Rectification (LR) and
Trust-anchored Reasoning (TR), together with Best-of-$N$ (BoN) as a
test-time scaling baseline. ``w/o LR \& TR'' corresponds to the vanilla
model without either component. Lower EHR is better, and the best results
for each backbone are bolded.
}
\label{tab:ablation_tts}

\begin{tabular}{lccccccccc}

\specialrule{1.2pt}{0pt}{0pt}
\noalign{\vskip 2pt}

\multirow{2}{*}{\textbf{Variant}}
& \multicolumn{9}{c}{\textbf{Emotion Hallucination Rate ($\downarrow$)}} \\

\cmidrule(lr){2-10}

& \textbf{Expression}
& \textbf{Action}
& \textbf{Audio}
& \textbf{Instinct}
& \textbf{Logic}
& \textbf{Conclusion}
& \textbf{Perception}
& \textbf{Psychology}
& \textbf{Total} \\

\midrule


AffectGPT
& 37.65 & 55.55 & 53.03
& 34.93 & 25.00 & 22.22
& 47.55 & 25.24 & 32.43 \\

\quad w/o TR
& 27.75 & 52.56 & 48.26
& 28.97 & 22.61 & 21.22
& 41.17 & 23.04 & 29.02 \\

\quad  w/o LR
& 21.80 & 41.94 & 36.30
& 26.12 & 23.71 & 21.19
& 30.96 & 23.09 & 25.29 \\

\quad  + BoN
& 32.08 & 56.03 & 51.93
& 31.65 & 22.26 & 21.35
& 45.57 & 23.41 & 30.78 \\

\rowcolor{tabhighlight}
\quad \textbf{+ Ours}
& \textbf{13.81}
& \textbf{39.31}
& \textbf{34.54}
& \textbf{25.70}
& \textbf{19.80}
& \textbf{21.17}
& \textbf{28.14}
& \textbf{22.50}
& \textbf{23.52} \\

\midrule

Qwen2.5-Omni
& 6.45 & 10.89 & 24.80
& 19.11 & 19.01 & 12.55
& 14.02 & 16.97 & 16.10 \\

\quad w/o TR
& 5.81 & 10.09 & 17.51
& 16.18 & 18.91 & 12.27
& 10.20 & 15.81 & 14.46 \\

\quad w/o LR
& 6.41 & 10.79 & 23.64
& 16.77 & 18.65 & 12.18
& 11.90 & 15.91 & 14.76 \\

\quad + BoN
& 5.64
& \textbf{7.05}
& 18.18
& 15.67
& \textbf{17.80}
& 15.67
& \textbf{8.47}
& 16.72
& 13.26 \\

\rowcolor{tabhighlight}
\quad \textbf{+ Ours}
& \textbf{4.31}
& 9.84
& \textbf{15.25}
& \textbf{12.84}
& 18.54
& \textbf{11.93}
& 9.14
& \textbf{14.28}
& \textbf{12.83} \\

\midrule

R1-Omni
& 25.38 & 59.23 & 67.61
& 50.89 & 28.99 & 46.01
& 42.16 & 40.62 & 41.28 \\

\quad w/o TR
& 19.60 & 52.97 & 50.72
& 45.80 & 28.74 & 42.28
& 33.38 & 38.12 & 35.57 \\

\quad w/o LR
& 23.00 & 56.74 & 64.59
& 48.65 & 23.29 & 45.02
& 38.46 & 37.85 & 38.12 \\

\quad + BoN
& 26.69 & 56.97 & 59.56
& 52.68 & 22.55 & 44.32
& 41.58 & 38.85 & 39.81 \\

\rowcolor{tabhighlight}
\quad \textbf{+ Ours}
& \textbf{14.09}
& \textbf{51.84}
& \textbf{41.49}
& \textbf{43.18}
& \textbf{17.49}
& \textbf{39.55}
& \textbf{30.10}
& \textbf{32.70}
& \textbf{31.47} \\

\specialrule{1.2pt}{0pt}{0pt}

\end{tabular}
\end{table*}
\begin{figure}[t]
    \centering
    \includegraphics[width=\linewidth]{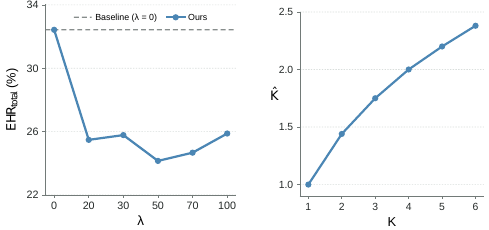}
    \caption{ \textbf{Hyperparameter analysis of \(\lambda\) and \(K\).} Performance peaks at \(\lambda=50\), and we set \(K=3\) to balance rectification effectiveness and efficiency. 
    }
    \label{fig:mitigate_analysis}
\end{figure}

\subsection{Emotion Hallucination Mitigation}

\noindent\textbf{Performance Comparison.}
Table~\ref{tab:overall_results} compares our method with VCD, ICD, M3ID, and EmotionHallucer across different backbones and datasets.
We make three observations.
(1) Our method is the only approach that reduces all six \texttt{EHR} facets in every setting.
In contrast, the baselines often improve some facets while degrading others.
For example, VCD increases \texttt{EHR}$_{per}$ from 42.16\% to 48.71\% and \texttt{EHR}$_{psy}$ from 40.62\% to 41.70\% on R1-Omni.
This suggests that object-centric mitigation may not transfer well to emotion hallucination, where the relevant evidence spans multiple modalities and cognitive facets.
ICD, M3ID, and EmotionHallucer also show facet-wise variations across backbones.
In particular, EmotionHallucer performs better on models with stronger instruction-following ability, such as Qwen2.5-Omni, but is less stable on other backbones.
(2) Our method demonstrates cross-dataset generalization.
On AffectGPT, our method reduces \texttt{EHR}$_{total}$ by 8.91, 9.89, and 11.85 percentage points on OV-MERD+, CMU-MOSI, and CH-SIMS, respectively.
It also reduces both perception and reasoning hallucinations on all three datasets, while the baselines show smaller or less consistent gains.
The results on CMU-MOSI and CH-SIMS further show that the method remains effective across English and Chinese datasets.
(3) Hallucination reduction does not sacrifice emotion prediction performance.
Our method improves F1 from 52.22 to 55.23 on AffectGPT and from 42.88 to 44.24 on R1-Omni.
On CMU-MOSI and CH-SIMS, F1 also increases from 77.77 to 78.41 and from 79.99 to 82.41, respectively.
Recall shows the same trend, including an increase from 53.00 to 58.45 on AffectGPT.
This suggests that suppressing hallucinated reasoning does not make the model overly conservative, but can preserve emotion-related evidence during prediction.

\begin{table}[t]
\centering
\caption{Wall-clock latency and token-cost under different maximum rectification iteration $K$.}
\setlength{\tabcolsep}{12pt}
\begin{tabular}{ccccc}
\specialrule{1.2pt}{0pt}{0pt}
\textbf{$K$} & \textbf{In-Token} & \textbf{Out-Token} & \textbf{Latency (s)} & \textbf{$\hat{K}$} \\
\midrule
1 & 940.82  & 137.39 & 79.80  & 1.00 \\
2 & 1303.95 & 214.54 & 124.43 & 1.44 \\
3 & 1559.42 & 271.70 & 157.44 & 1.75 \\
4 & 1757.64 & 316.26 & 183.41 & 2.00 \\
5 & 1926.14 & 352.62 & 204.99 & 2.20 \\
6 & 2071.80 & 384.87 & 224.07 & 2.38 \\
\specialrule{1.2pt}{0pt}{0pt}
\end{tabular}
\label{tab:token_cost}
\end{table}

\noindent\textbf{Ablation Study.} \textit{\textbf{(1) Main Module.}} Table~\ref{tab:ablation_tts} presents the ablation results of Logit Rectification (LR) and Trust-anchored Reasoning (TR) on AffectGPT. Compared to the baseline, introducing LR reduces the total hallucination rate from 32.43\% to 29.02\%, demonstrating that token-level logit rectification effectively suppresses hallucinated tokens during decoding. When only TR is enabled, the hallucination rate further decreases to 25.29\%, showing a larger improvement than LR alone. This suggests that progressive sentence verification plays a more critical role than post-hoc token correction in preventing error snowballing in long-form generation. Finally, combining LR and TR yields the best performance, consistently outperforming all variants across evaluation facets. \textit{\textbf{(2) Test-time Scaling Alternative.}} 
Table \ref{tab:ablation_tts} compares \texttt{HMER} with the Best-of-$N$ (BoN) sampling strategy using evaluator-produced \texttt{EHR} as the selection score. BoN reduces \texttt{EHR} compared with vanilla baselines, supporting the evaluator's ability to rank reasoning quality during test-time scaling~\cite{venktesh2025trust}. However, BoN relies on output diversity: it performs well on models that generate diverse reasoning paths (\eg, Qwen2.5-Omni), but is less effective when sampled responses converge to similar erroneous reasoning patterns. In contrast, \texttt{HMER} intervenes during generation to suppress hallucination pathways and guide the model toward grounded reasoning, consistently achieving lower $\texttt{EHR}_{psy}$ across all three models.

\noindent\textbf{Hyperparameter Analysis.}
The hallucination penalty coefficient $\lambda$ and the maximum rectification iteration $K$ are two key hyperparameters in our framework.
As shown in Fig.~\ref{fig:mitigate_analysis}, setting $\lambda$=0 disables hallucination suppression, resulting in a substantially higher hallucination rate. Introducing a moderate penalty consistently reduces hallucinations, with the best performance observed around $\lambda$=50. 
Besides, we examine the average effective round $\hat{K}$ under different $K$. $\hat{K}$ is defined as the iteration where a sentence is verified as hallucination-free. As shown in Fig.~\ref{fig:mitigate_analysis}, $\hat{K}$ increases with $K$, indicating that iterative refinement improves sentence reliability. However, the gains gradually saturates, while inference latency grows steadily. Considering the trade-off between reliability and efficiency, we adopt $K=3$ in all experiments.


.\begin{figure*}[!t]
    \centering
    \includegraphics[width=\textwidth]{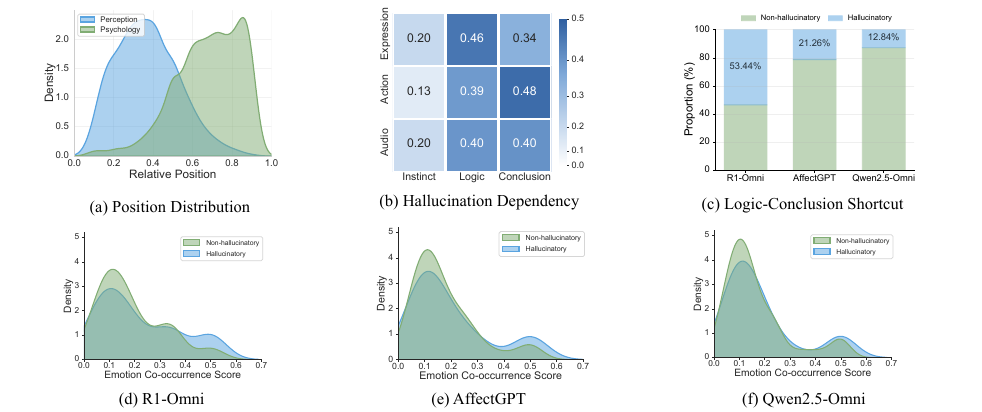}
    \caption{\textbf{Hallucination mechanism analysis}. (a) Positional distribution of perceptual and reasoning elements in generated emotion interpretations. (b) Co-occurrence matrix between different hallucination types. (c) Dependency between logic and conclusion reliability across different MLLMs. (d) (e) (f) Distributions of emotion co-occurrence scores for hallucinated and non-hallucinated pairs on R1-Omni, AffectGPT, and Qwen2.5-Omni.
}
    \label{fig:mechanism}
\end{figure*}

\begin{figure*}[!t]
  \centering
  \includegraphics[width=\textwidth]{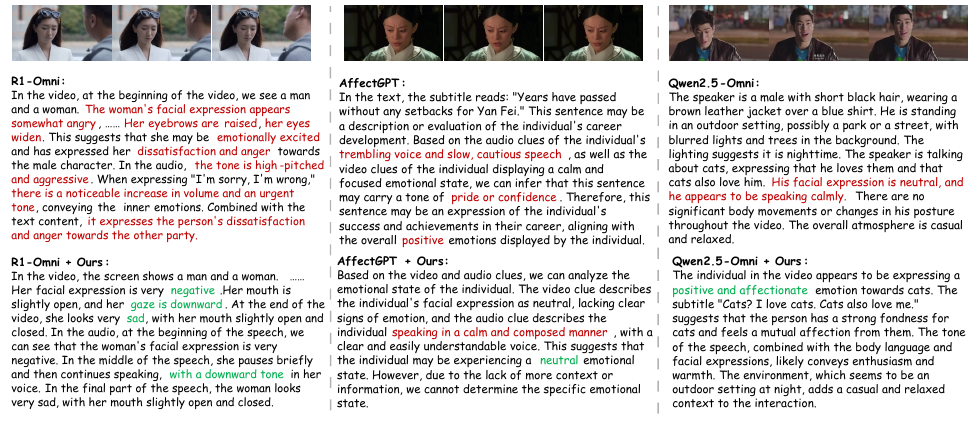} 
  \caption{\textbf{Case study} on R1-Omni, AffectGPT, and Qwen2.5-Omni. The red and green text represent hallucinatory words and correct words, respectively. The cases show that our method effectively rectifies incoherent emotional inference. 
  }
  \label{fig:results_sample}
\end{figure*}

\noindent\textbf{Wall-Clock Latency and Token-Cost}
To evaluate the computational overhead of iterative rectification, we analyze token consumption and wall-clock latency under different maximum rectification iterations $K$, as shown in Table~\ref{tab:token_cost}. As $K$ increases, both input/output token usage and end-to-end latency grow accordingly, since each additional iteration introduces a new evaluation–rectification cycle. However, the average effective iteration $\hat{K}$ (the round at which generated claims are verified as hallucination-free) remains consistently lower than the maximum $K$. The early stopping mechanism keeps the actual inference cost relatively controlled.

\subsection{Hallucination Mechanism Analysis}
\noindent\textbf{Structured Dependency in Emotion Reasoning.}
To investigate the structural origin of emotion hallucination, we first analyze the positional distribution of perceptual and reasoning elements in Fig.~\ref{fig:mechanism} (a). It reveals a natural transition from observation to abstraction in long-form generation. In this context, Fig.~\ref{fig:mechanism} (b) further examines cross-type hallucination dependencies. It can be found that early perceptual hallucinations tend to influence subsequent logical inference and final conclusions. Such dependency patterns provide empirical evidence for the snowballing phenomenon observed in affective reasoning.

\noindent\textbf{Shortcut Reliance in Conclusion Formation.}
We next analyze the relationship between logical reliability and conclusion reliability across models (Fig.~\ref{fig:mechanism} (c)). We measure the proportion of correct \textit{Logic} but incorrect \textit{Conclusion} in test set. When logical reasoning is verified as reliable, the probability of conclusion hallucination remains relatively low in the general-purpose model Qwen2.5-Omni (12.84\%), but is higher in emotion-specialized models AffectGPT (21.26\%) and R1-Omni (53.44\%). This gap may reflect a redistribution of model capacity after emotion-specific fine-tuning. The model may allocate greater emphasis to emotion-consistent surface patterns, potentially reducing the implicit constraint between intermediate reasoning and final conclusions.

\noindent\textbf{Emotion Co-occurrence Bias.}
Finally, we analyze emotion word pair co-occurrence patterns in Fig.~\ref{fig:mechanism} (d). Across Qwen2.5-Omni, AffectGPT, and R1-Omni, hallucinated pairs tend to concentrate in high co-occurrence pairs compared to non-hallucinated pairs. In particular, the lightweight R1-Omni exhibits the largest proportion of high co-occurrence hallucinated pairs. These findings indicate that strong statistical associations between emotion terms can bias generation toward overconfident yet insufficiently grounded predictions.

\noindent \textbf{Case study.}
Fig.~\ref{fig:results_sample} presents qualitative examples of hallucination mitigation across three backbones. 
(1) Our method effectively rectifies different facets of emotion understanding. For instance, it corrects audio hallucinations (\eg, ``increased volume'' $\rightarrow$ ``downward tone'' on R1-Omni) and instinct hallucinations (\eg, ``neutral'' $\rightarrow$ ``positive and affectionate'' on Qwen2.5-Omni). 
(2) Hallucinated perception can introduce modality conflicts that lead to contradictory reasoning. On the R1-Omni, hallucinated ``eyebrows raised'' suggests ``excited,'' while an ``aggressive tone'' implies ``anger''. 
By grounding perceptual evidence (\eg, ``gaze is downward'' and ``downward tone'') as trusted anchors, our method restores a coherent reasoning chain and infers the correct ``sad'' emotion. 
(3) The cases also reveal the limitation of the rectification process: the model may occasionally rely on conservative reasoning anchors, producing cautious or repetitive interpretations (\eg, ``cannot determine'' on AffectGPT). This reflects a trade-off between hallucination suppression and exploration during generation.

\section{Conclusion}
In this paper, we introduce \texttt{EHR}, a psychologically grounded protocol for evaluating emotion hallucinations across multimodal perception and affective reasoning. Experiments across diverse MLLMs reveal that these failures are prevalent and can propagate along open-ended reasoning trajectories. To mitigate them, we propose \texttt{HMER}, which converts localized evaluator feedback into trajectory-level logit rectification and maintains low-risk reasoning states as recurrent anchors. \texttt{HMER} consistently reduces hallucination rates across representative model backbones while preserving emotion prediction quality. Our findings provide a systematic basis for evaluating and regulating the reliability of open-ended multimodal emotion reasoning.

\bibliographystyle{IEEEtran}
\bibliography{main}

\end{document}